\documentclass[a4paper,fleqn,usenatbib]{mnras}
\usepackage[T1]{fontenc}
\usepackage{ae,aecompl}

\usepackage[normalem]{ulem}
\usepackage{amsmath}
\usepackage{graphicx}	
\usepackage{lscape}
\usepackage{indentfirst}
\usepackage{enumitem}

\usepackage{amssymb}
\usepackage{color}
\usepackage[flushleft]{threeparttable}

\newcommand {\bc}{\begin {center}}
\newcommand {\ec}{\end {center}}
\newcommand {\be}{\begin {equation}}
\newcommand {\ee}{\end {equation}}
\newcommand {\beq}{\begin {eqnarray}}
\newcommand {\eeq}{\end {eqnarray}}

\newcommand {\ergs}{{\rm erg\ \rm s^{-1}}}
\newcommand {\comment}[1]{}

\def\lbar {\lambda\hskip-5pt\raise3pt\hbox {--}}
\def\lbr {\lambda\raise2pt\hbox {\hskip-4pt{\scriptsize --}}_\C}

\renewcommand{\d}{{\rm d}}

\def\ff {{\rm f}}
\def\ii {{\rm i}}

\def\lum{erg s$^{-1}$}

\renewcommand{\d}{{\rm d}}

\title[Spectra of XRPs at low luminosities]
{Spectrum formation in X-ray pulsars at very low mass accretion rate: Monte-Carlo approach}
\author[A.~A.~Mushtukov et al.] 
{Alexander~A.~Mushtukov,$^{1,2}$\thanks{E-mail: al.mushtukov@gmail.com (AAM)}  
Valery~F.~Suleimanov,$^{3,4,2}$
Sergey~S.~Tsygankov$^{5,2}$ 
\newauthor
and Simon Portegies Zwart$^{1}$ \\ 
$^1$ Leiden Observatory, Leiden University, NL-2300RA Leiden, The Netherlands \\
$^2$ Space Research Institute of the Russian Academy of Sciences, Profsoyuznaya Str. 84/32, Moscow 117997, Russia \\
$^3$ Institut f\"ur Astronomie und Astrophysik, Universit\"at T\"ubingen, 
    Sand 1, D-72076 T\"ubingen, Germany \\
$^4$  Kazan Federal University, Kremlevskaya str. 18, Kazan 42008, Russia\\
 $^5$ Department of Physics and Astronomy,  FI-20014 University of Turku, Finland
} 

\pubyear{2021}

\begin{document}
\label{firstpage}
\pagerange{\pageref{firstpage}--\pageref{lastpage}}
\maketitle

\begin{abstract}
It has been recently discovered that the transition of X-ray pulsars to the low luminosity state ($L\lesssim10^{35}\,\ergs$) is accompanied by a dramatic spectral change. Namely, the typical power-law-like spectrum with high energy cutoff transforms into a two-component structure with a possible cyclotron absorption feature on top of it.
It was proposed that these spectral characteristics can be explained qualitatively by the emission of cyclotron photons in the atmosphere of the neutron star caused by collisional excitation of electrons to upper Landau levels and further comptonization of the photons by electron gas. The latter is expected to be overheated in a thin top layer of the atmosphere.
In this paper, we perform Monte Carlo simulations of the radiative transfer in the atmosphere of an accreting neutron star while accounting for a resonant scattering of polarized X-ray photons by thermally distributed electrons.
The spectral shape is shown to be strongly polarization-dependent in soft X-rays ($\lesssim 10\,{\rm keV}$) and near the cyclotron scattering feature.
The results of our numerical simulations are tested against the observational data of X-ray pulsar A~0535+262 in the low luminosity state.
We show that the spectral shape of the pulsar can be reproduced by the proposed theoretical model. The applications of the discovery to the observational studies of accreting neutron stars are discussed.
\end{abstract}



\section{Introduction}
\label{sec:Intro}

Classical X-ray pulsars (XRPs) are accreting strongly magnetized neutron stars (NSs) orbiting around optical stellar companions \citep[for a review see e.g.,][]{2015A&ARv..23....2W}. 
One of the richest families of XRPs are systems with Be star as a companion \citep[BeXRPs;][]{2011Ap&SS.332....1R}. These systems exhibit strong transient activity that allows us to study radiation's interaction with a matter under conditions of extremely strong magnetic fields over a wide range of mass accretion rates covering up to six orders of magnitude \citep{2020MNRAS.491.1857D}.

From an observational point of view, spectra of XRPs at high luminosities ($>10^{36}$~\lum) have similar shapes which can be well fitted by a power-law with an exponential cutoff at high energies \citep[see, e.g., ][]{1989PASJ...41....1N, 2005AstL...31..729F}. 
However, it was recently discovered that decrease
of the observed luminosity below this value is accompanied by the dramatic changes of the energy spectra {in several XRPs} \citep{2019MNRAS.483L.144T,2019MNRAS.487L..30T,2021arXiv210110834D,2021arXiv210305728L}, pointing to varied physical and geometrical properties of the emission region. 

\begin{figure*}
\centering 
\includegraphics[width=14.cm, angle =0]{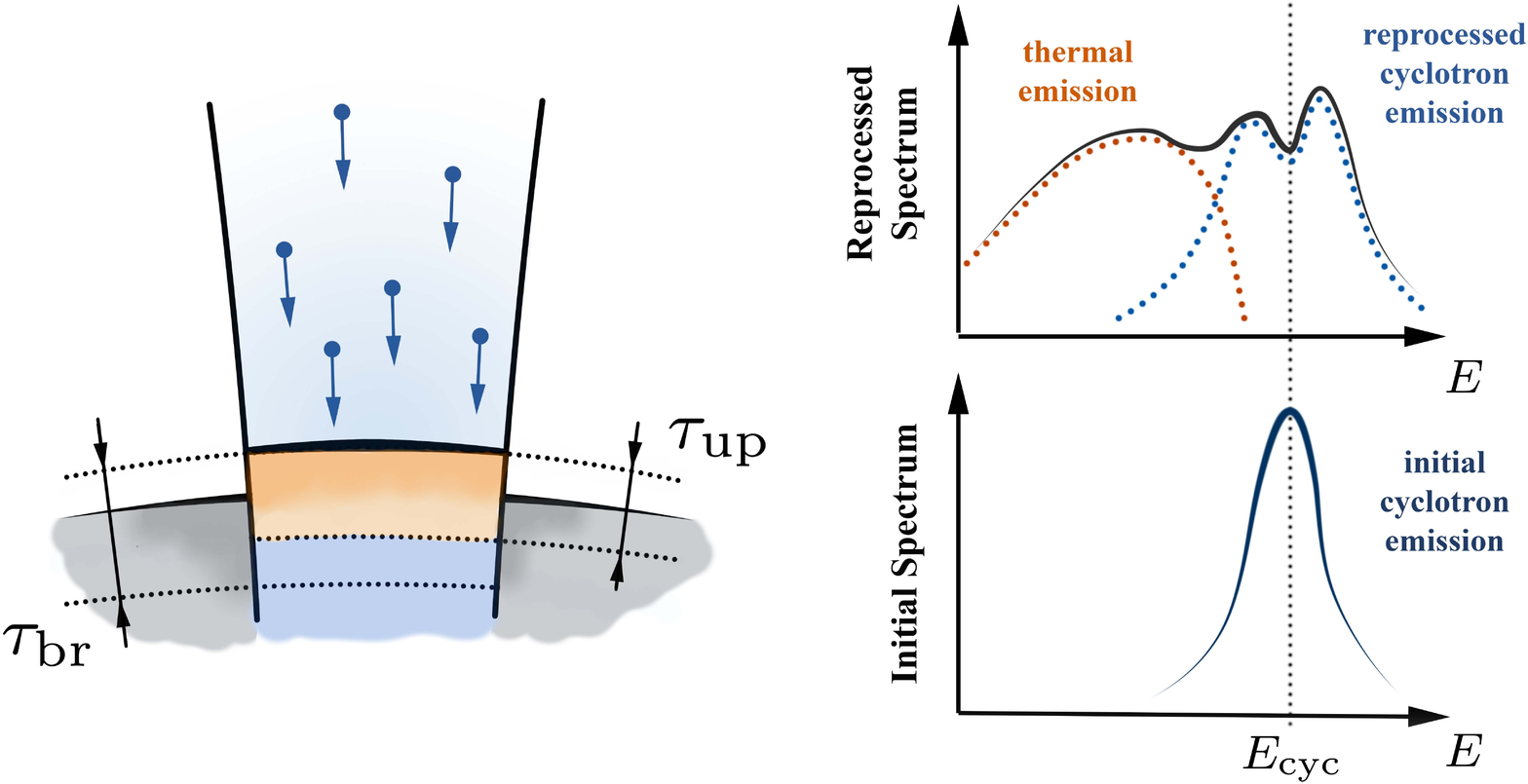} 
\caption{
Schematic picture of the theoretical model.
The X-ray energy spectrum is originated from the atmosphere of a NS with the upper layer overheated by low-level accretion.
The accretion flow is stopped in the atmosphere due to collisions.
Collisions result in excitation of electrons to upper Landau levels.
The following radiative de-excitation of electrons produces cyclotron photons.
The cyclotron photons are partly reprocessed by magnetic Compton scattering and partially absorbed in the atmosphere.
The reprocessed photons form a high-energy component of the spectrum, while the absorbed energy is released in thermal emission and forms a low-energy part of the spectrum.
}
\label{pic:000}
\end{figure*}

{In particular, originally it was shown for the transient XRPs, GX~304$-$1 and A~0535+26, that once their} luminosity drops down to $~10^{34}-10^{35}$~\lum, the "canonical" spectral shape of their emission undergoes a dramatic transition into a two-component structure consisting of two humps peaking at $\sim5-7$~keV and $\sim30-50$~keV \citep{2019MNRAS.483L.144T,2019MNRAS.487L..30T,2021arXiv210305728L}. 
The only other source with a similar spectral structure observed earlier is the persistent low-luminosity XRP X Persei \citep{2012A&A...540L...1D}, which, however, never showed spectral transitions as a function of luminosity.
In the case of A~0535+26, a cyclotron absorption feature was also detected on top of the high energy component of the spectrum. The appearance of the high-energy spectral component was interpreted by the recombination of electrons collisionally excited to the upper Landau levels in the heated layers of the NS atmosphere (see Fig.\,\ref{pic:000}, \citealt{2019MNRAS.487L..30T}). 
However, no quantitative description had been done.

Dramatic spectral changes in XRPs at very low mass accretion rates can shed light on different physical processes responsible for the spectra formation at different luminosities. Depending on the luminosity, one can distinguish different regimes of accretion, i.e., the interaction of radiation with accreting plasma:\\
(i) At high mass accretion rates, the luminosity and local X-ray energy flux are high enough to stop accretion flow above NS surface in radiation-dominated shock \citep{1982STIN...8325627L}.
Below the shock region, the matter slowly settles down to the stellar surface and forms the so-called accretion column confined by a strong magnetic field \citep{1975PASJ...27..311I,1976MNRAS.175..395B,1981A&A....93..255W,2015MNRAS.454.2539M,2021MNRAS.501..564G}.
Because of magnetic field confinement and opacity reduced in a strong magnetic field \citep{1979PhRvD..19.2868H}, the accretion column can survive under the extreme conditions of high radiation pressure. 
The minimal accretion luminosity, which is sufficient to stop the matter in radiation pressure dominated shock, is called the critical luminosity $L_{\rm crit}$ \citep{1976MNRAS.175..395B}.
The critical luminosity is of the order of $10^{37}\,\ergs$, and depends on the geometry of accretion flow and magnetic field strength at the NS surface \citep{2012A&A...544A.123B,2015MNRAS.447.1847M}.\\
(ii) At low mass accretion rates (sub-critical regime, when $L<L_{\rm crit}$), the radiation pressure is insufficient to stop accretion flow above NS surface, and final braking of the flow happens in the atmosphere of a NS due to collisions between particles.
Sub-critical accretion results in hot spots at the stellar surface located close to the NS's magnetic poles.
Even in the sub-critical regime, the accretion flow dynamics can be affected by radiation pressure, and X-ray spectra can be shaped by the interaction of photons with the accreting material (see, e.g., \citealt{2007ApJ...654..435B,2015MNRAS.454.2714M}).
Only in the case of very low mass accretion rates ($\dot{M}\lesssim 10^{15}\,{\rm g\,s^{-1}}$) the influence of interaction between accreting matter and radiation above the NS surface is negligible. 
In this regime, almost all energy of accretion flow is released in the geometrically thin NS atmosphere, where the accreted protons transfer their kinetic energy to the plasma heating. 
In this case, the geometry of the emitting region is elementary, and the plane-parallel approximation for the particle heated atmosphere is a reasonable approximation for a complete description of spectra formation.

In this work, based on the Monte-Carlo simulations of radiative transfer in a strong magnetic field, we propose the first simple two-slab model describing the spectral formation in XRPs at very low mass accretion rates. 
In Section~\ref{sec:Model}, we introduce our model set up and the basic assumptions of radiative transfer. 
The basic details of our numerical code are given in Section~\ref{sec:MonteCarlo}.
Results of numerical simulations and their comparison with the observational data are given in Section~\ref{sec:NumRes}.
Summarizing our results in Section~\ref{sec:sum}, we propose applications to the observational studies and argue that the proposed concept of spectra formation at extremely low state of accretion luminosity can be used in detailed diagnostics of ``propeller" state in accretion and be a base of alternative estimations of NS magnetic field strength.

\section{Model set up}
\label{sec:Model}

\subsection{Geometry of emitting region}

Sub-critical mass accretion rates (see e.g. \citealt{1976MNRAS.175..395B,2015MNRAS.447.1847M}) result in the simplest geometry of emitting regions: the X-ray photons are produced by hot spots at the NS surface.
If the mass accretion rate is well below the critical luminosity value $L\ll L_{\rm crit}$, the influence of X-ray energy flux on the dynamics of the accretion process is insignificant \citep{2015MNRAS.454.2714M}.

In the case of magnetic field dominated by dipole component, the area of the spots at the NS surface can be roughly estimated as
\begin{eqnarray}
S \approx 3 \times 10^9\, \Lambda^{-7/8}\, m^{-13/20}\, R_6^{19/10}\, B_{12}^{-1/2}\, L_{37}^{2/5}\quad \mbox{cm}^2, 
\end{eqnarray}
where $\Lambda$ is a constant which depends on accretion flow geometry: $\Lambda<1$ for the case of accretion through the disc with $\Lambda=0.5$ being a commonly used value (see e.g. \citealt{1978ApJ...223L..83G,2014EPJWC..6401001L}), $m$ is the NS mass measured in units of solar masses $M_\odot$, $R_6$ is the NS radius measured in $10^6\,{\rm cm}$, $B_{12}$ is the magnetic field strength at the pole of a NS measured in units of $10^{12}\,{\rm G}$, and $L_{37}$ is the accretion luminosity of XRP measured in units of $10^{37}\,\ergs$.
The corresponding effective temperature of a hot spots can be estimated as
\beq\label{eq:Teff}
T_{\rm eff} = \frac{L}{2\sigma_{\rm SB} S} = 6.6 \, \Lambda^{7/32}\, m^{13/80}\, R_6^{-19/40}\, B_{12}^{1/8}\, L_{37}^{3/20} \quad \mbox{keV},
\eeq
where $\sigma_{\rm SB}$ is the Stefan–Boltzmann constant.

In the case of accretion on weakly magnetized NSs, the protons of accretion flow brake in a stellar atmosphere due to Coulomb interactions with atmospheric electrons (see Chapter 3 in \citealt{2002apa..book.....F}). 
In the case of highly magnetized atmospheres, the physical picture is similar, but a significant fraction of proton kinetic energy turns into excitation of electrons to upper Landau levels \citep{1995ApJ...438L..99N}.
Further de-excitation of electrons results in production of cyclotron photons at energy $E_{\rm cyc}\approx 11.6\,B_{12}\,{\rm keV}$ (it is a case if $B\lesssim 10^{13}\,{\rm G}$, see e.g. \citealt{1987ApJ...319..939H}).
It is expected that the sources of cyclotron photons decay exponentially with the optical depth $\propto e^{-\tau/\tau_{\rm br}}$, where $\tau_{\rm br}$ is a typical optical depth due to Thomson scattering, where the accretion flow is braked due to collisions.
The total number of cyclotron photons emitted in the atmosphere per ${\rm cm}^2$ in a ${\rm sec}$ due to accretion flow braking can be estimated as 
\beq 
\frac{1}{S}\frac{\d N_{\rm cyc}}{\d t}\approx
6.2\times 10^{35}\frac{L_{37}}{E_{\rm cyc,keV}S_{10}}\quad
{\rm cm^{-2}s^{-1}},
\eeq 
where $S_{10}$ is the area of a hot spot at the NS surface in units of $10^{10}\,{\rm cm}$.

A temperature structure of the accretion heated weakly magnetized atmosphere could be divided into two qualitatively different layers. 
The most of energy is released in deep optically thick layers (if $\tau_{\rm br} > 1$) where the plasma mass density is high enough and cooling due to free-free emission can compensate the heating due to braking of the accretion flow. 
The deep heated layer is close to being isothermal, with the temperature close to the effective one (\ref{eq:Teff}). 
However, the upper rarefied layers of the atmosphere have too low density to be able to cool with free-free emission.
As a result, the upper layers are cooled by Compton scattering and overheated up to tens and even a hundred keV (see, e.g., \citealt{2018A&A...619A.114S}). 
We assume that the highly magnetized accretion heated atmospheres have a similar qualitatively structure.
Thus, we consider a plane-parallel magnetized semi-infinite atmosphere with a thin overheated upper layer and an isothermal deeper layer (see Fig.\,\ref{pic:000}). 
The temperatures of both parts are free parameters of our model, as well as the optical thickness of the overheated slab. 
The magnetic field is taken to be perpendicular to the NS surface, which is a good approximation for the regions located close to the magnetic poles of a NS.
The resulting spectrum is computed by solving the radiation transfer equation using the Monte-Carlo method.

\subsection{Radiative transfer}
\label{sec:RadTransfer}

Both opacity and refractive index are dependent on photon polarization in a strongly magnetized plasma and vacuum.
As a result, the solution of the radiative transfer problem has to account for the polarization of photons.
In the general case, the polarization of X-ray photons can be described in terms of four Stokes parameters. 
At the same time, the radiative transfer equation turns into a set of four equations: one for each Stokes parameter.
Because the plasma in a strong magnetic field is anisotropic and birefringent \citep{1970pewp.book.....G} the polarization of X-ray photon can vary along its trajectory. 
It makes the description of polarised radiative transfer by the Stokes parameters even more complicated. 
However, any X-ray photon can be represented as  a linear composition of two orthogonal normal modes, which conserve their polarisation state along their trajectories (see, e.g., \citealt{1977ewcp.book.....Z}). 
The linear composition of the normal modes changes its polarisation due to the difference in the phase velocity of the modes.
If the difference between the phase velocities of normal modes is large enough, the radiative transfer problem reduces and can be effectively solved for two normal modes, i.e., the radiation can be described by specific intensities in two normal polarization modes $I_E^{(s)}$, where $s=1\,\,(s=2)$ corresponds to X-mode (O-mode).

The ellipticity of normal modes is determined by
plasma conditions (temperature, mass density and chemical composition, see e.g., \citealt{1970pewp.book.....G,1980PlPh...22..639K}) and vacuum polarization (see \citealt{2006RPPh...69.2631H} for review).
In this paper, we neglect the effects of vacuum polarization (see Appendix\,\ref{App:vac_pol}), assume that
the dielectric tensor in the atmosphere coincides with the dielectric tensor of cold plasma, 
the normal modes are transverse and use the approximated ellipticity of the modes affected by plasma effects only:
\beq\label{eq:NormWavEll}
K_s(E,\theta) &\equiv& -i\left(\frac{E_x}{E_y}\right) \\
&=& \frac{2\cos\theta}{\frac{E_{\rm cyc}}{E}\sin^2\theta -(-1)^{s} \sqrt{\frac{E^2_{\rm cyc}}{E^2}\sin^4\theta +4\cos^2\theta} }, 
\nonumber
\eeq
where $E_x$ ($E_y$) is the photon's electric field component along (perpendicular to) the $\mathbf{k}-\mathbf{B}$ plane, where $\mathbf{k}$ is the vector of photon momentum and $\mathbf{B}$ is the vector of external magnetic field, and $\theta$ is the angle between the photon momentum and $B$-field direction \citep{1973ZhETF..65..102G}. 

The main processes of interaction between radiation and matter in our model are 
\begin{itemize}
\item magnetic Compton scattering, 
\item cyclotron absorption,
\item bremsstrahlung affected by a strong magnetic field.
\end{itemize}
The radiative transfer equation can be written as
\beq\label{eq:RTE}
\cos\theta\frac{\d I_E^{(s)}(\Omega)}{\d \tau_{\rm T}}&=&\frac{\alpha_{\rm abs}^{(s)}(\Omega)}{\alpha_{\rm T}}I_E^{(s)}(\Omega)\\
&-&\sum\limits_{j=1}^{2}\int\limits_{0}^{\infty}\d E'\int\limits_{(4\pi)}\d\Omega'\nonumber\\
&&\times\left[R_{sj}(E,\Omega| E',\Omega') I_{E'}^{(j)}(\Omega') \right. \nonumber \\
&&- \left. R_{js}(E',\Omega' | E,\Omega)I_{E}^{(s)}(\Omega)\right] \nonumber \\
&-& \frac{\alpha_{\rm abs}^{(s)}(\Omega)}{\alpha_{\rm T}}\frac{B_E}{2} - S_{\rm ini}^{(s)}(E,\Omega), \nonumber
\eeq
where 
$\tau_{\rm T}$ is the optical depth due to non-magnetic Thomson scattering, $\alpha_{\rm abs}^{(s)}$ is the coefficient of true absorption (due to free-free and cyclotron mechanisms), $\alpha_{\rm T}$ is the absorption coefficient due to non-magnetic Thomson scattering, 
$B_E(T)$ is the Planck function.
The redistribution function $R_{s\,s'}(E,\Omega | E',\Omega')$ describes redistribution of X-ray photons over energy, momentum and polarization state, and 
is related to the probability of photon transition from energy $E'$, direction given by solid angle $\Omega'$ and polarisation state $s'$ to energy $E$, direction $\Omega$ and polarization state $s$ due to magnetic Compton scattering.
The radiative transfer equation (\ref{eq:RTE}) neglects non-linear effects like induced scattering because they are expected to be negligible in low-luminosity states.

The first term in the right-hand side of equation (\ref{eq:RTE}) describes absorption due to bremsstrahlung and cyclotron mechanism. 
With the second term, we account for the photon redistribution due to Compton scattering. 
The third term introduces the thermal emission of photons, which is calculated in the assumption of local thermodynamic equilibrium (LTE).
The last term gives the primary sources of cyclotron emission, which are distributed exponentially in our particular case $S_{\rm ini}\propto e^{-\tau/\tau_{\rm br}}$.

\begin{figure*}
\centering 
\includegraphics[width=15.cm, angle =0]{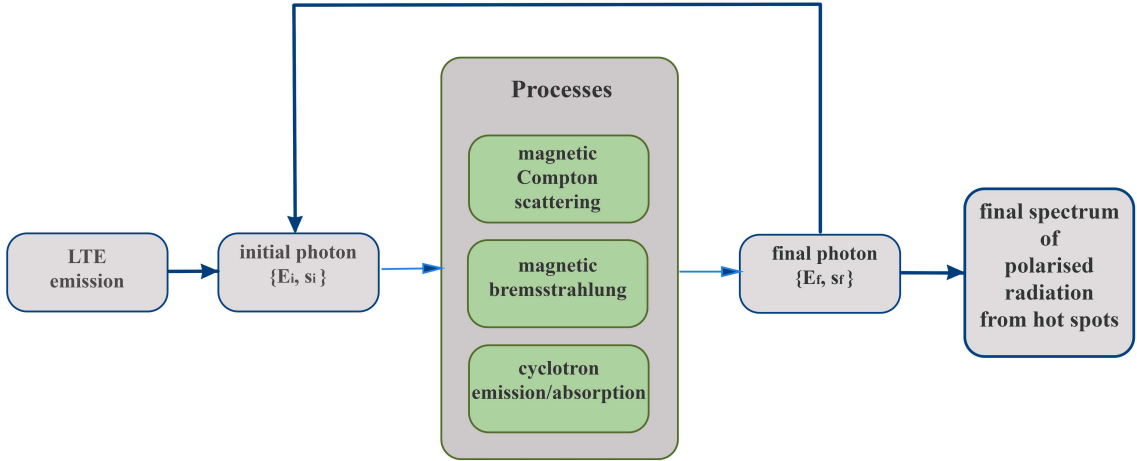} 
\caption{
The scheme of Monte Carlo simulations performed in the paper.  
}
\label{pic:NumSch}
\end{figure*}

\begin{figure*}
\centering 
\includegraphics[width=16.cm]{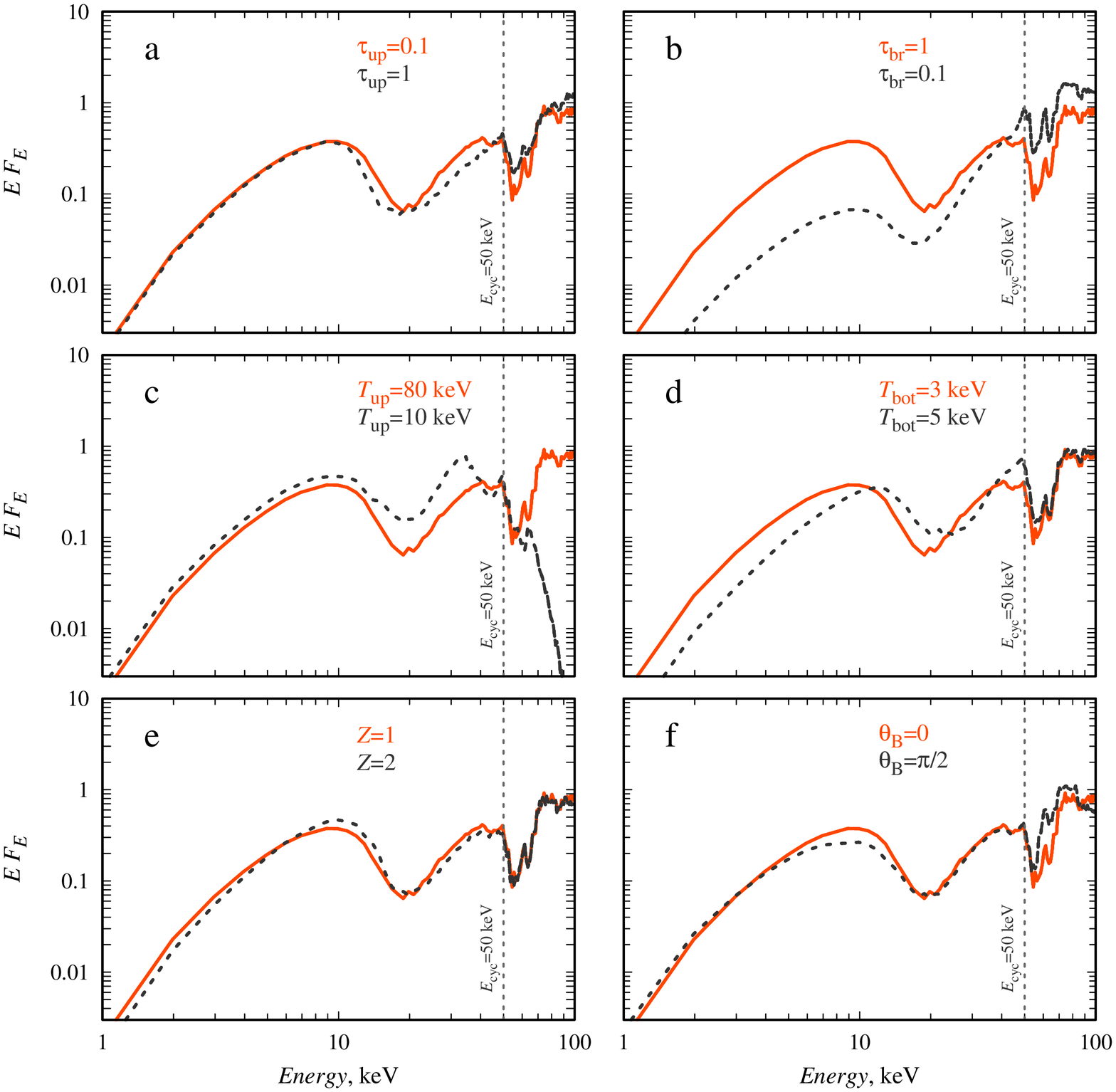} 
\caption{
$E\,F_E$ spectrum ({in the arbitrary units}) of X-ray radiation leaving the atmosphere of a NS at very low mass accretion rates.
The fiducial case is given by red solid line and corresponds to the case of atmosphere with describing by the following set of parameters: $E_{\rm cyc}=50\,{\rm keV}$, $T_{\rm bot}=3\,{\rm keV}$, $T_{\rm up}=80\,{\rm keV}$, $\tau_{\rm up}=0.1$, $\tau_{\rm br}=1$, $Z=1$.
Different panels show the influence of different parameters on the final X-ray energy spectrum.
(a) The effect of optical thickness $\tau_{\rm up}$ of the overheated upper layer.
(b) The influence of effective depth $\tau_{\rm br}$, where the accretion flow is stopped by collisions.
(c) The influence of the temperature $T_{\rm up}$ of the overheated upper layer. 
(d) The influence of the temperature $T_{\rm bot}$ of the atmosphere below the overheated upper layer.
(d) The influence of the chemical composition of the atmosphere.
(f) The influence of the angle $\theta_{\rm B}$ between the magnetic field lines and normal to the NS surface.
The run results from $2\times 10^6$ photons, leaving the atmosphere.
}
\label{pic:MC_comparison}
\end{figure*}

\subsubsection{Compton scattering}

The redistribution function due to Compton scattering in (\ref{eq:RTE}) is given by
\beq
R_{s_{\rm f},s_{\rm i}}(E_{\rm f},\Omega_{\rm f} | E_{\rm i},\Omega_{\rm i})=
\frac{1}{\sigma_{\rm T}}
\frac{\d\sigma_{s_{\rm f},s_{\rm i}}}{\d\Omega_{\rm f} \d E_{\rm f}}([f_{\rm e}(p_{b})],E_{\rm i},\Omega_{\rm i},\Omega_{\rm f}),
\eeq
where the double differential cross section of scattering by the electron gas with the electron distribution over the momentum $f_{\rm e}(p_{b})$ is
\beq
 \frac{\d\sigma_{s_{\rm f},s_{\rm i}}}{\d\Omega_{\rm f}\d E_{\rm f}}([f_{\rm e}(p_{b})],E_{\rm i},\Omega_{\rm i},E_{\rm f},\Omega_{\rm f})&=&
 \frac{\d\sigma_{s_{\rm f},s_{\rm i}}}{\d\Omega_{\rm f}}(p_{b}^*,E_{\rm i},\Omega_{\rm i},\Omega_{\rm f}) \nonumber \\
 &&\times
  \left( \frac{\d p_{b}^*}{\d E_{\rm f} } \right)
  f_{\rm e}(p_{b}^*), 
\eeq
the solid angles are related to the spherical coordinates describing the direction of photon motion $(\theta_{\ii,\ff},\varphi_{\ii,\ff})$ as
$\d\Omega_{\ii,\ff}=\sin\theta_{\ii,\ff}\d\theta_{\ii,\ff}\d\varphi_{\ii,\ff}$,
$p^*_{b}$ is the electron momentum required for the photon transition $\{E_{\rm i},\Omega_{\rm i}\}\longrightarrow \{E_{\rm f},\Omega_{\rm f}\}$,
\beq
\frac{\d\sigma_{s_{\rm f},s_{\rm i}}}{\d\Omega_{\rm f}}(p_{b},E_{\rm i},\Omega_{\rm i},\Omega_{\rm f})&=&
\frac{\d\sigma_{s_{\rm f},s_{\rm i}}}{\d\Omega_{\rm f}}(p_{b}=0,E^*_{\rm i},\theta^*_{\rm i},\varphi_{\rm i},\theta^*_{\rm f},\varphi_{\rm f}) \nonumber \\
&&\times\frac{(1-\beta^2)}{(1-\beta\cos\theta_{\rm f})^2},
\eeq
is the ordinary scattering cross section by the electron of momentum $p_{b}$ along magnetic field lines,
and 
\beq 
E^*_{\rm i}&=&E_{\rm i}\gamma (1-\beta\cos\theta_{\rm i}), \\
\cos\theta^*_{\rm i,f}&=&\frac{\cos\theta_{\rm i,f}+\beta}{1+\beta\cos\theta_{\rm i,f}} \\
\gamma&=&[1+p_b^2/(m_{\rm e}c)^2]^{1/2}=(1-\beta^2)^{-1/2}
\eeq 
account for the Doppler effect and aberration due to transition between different reference frames.

Compton scattering is affected by a strong magnetic field {in the atmosphere of a NS} \citep{1986ApJ...309..362D}. 
The scattering cross-section strongly depends on photon energy $E$, momentum, and polarization state.
The scattering of photons which energy is close to the cyclotron energy $E_{\rm cyc}\approx 11.6\,(B/10^{12}\,{\rm G})\,{\rm keV}$ becomes resonant.
In this paper, Compton scattering is considered in a non-relativistic manner (see e.g., \citealt{1979PhRvD..19.2868H}).
The differential cross sections of Compton scattering by electron at rest are expressed by the complex amplitudes of scattering $a^{\rm (p)}_{s_{\rm f}s_{\rm i}}$:
\beq 
\label{eq:Amp2CS}
\frac{\d\sigma_{s_{\rm f}s_{\rm i}}}{\d\Omega_{\rm f}}(p_{b}=0,E_{\rm i},\theta_{\rm i},\varphi_{\rm i},\theta_{\rm f},\varphi_{\rm f})
=\frac{3}{32\pi}\sigma_{\rm T}|a^{\rm (p)}_{s_{\rm f}s_{\rm i}}|^2,
\eeq
where $E_{\rm i}$ is the initial photon energy,
$s_{\rm i}$ and $s_{\rm f}$ denote the photon polarization state before and after the scattering respectively.
The complex amplitudes $a^{\rm (p)}_{s_{\rm f}s_{\rm i}}$ depend on exact expression for the polarization modes.
The amplitudes can be combined in a matrix
\beq
\widehat{a}^{\rm (p)}=
\left(\begin {array}{cc} a^{\rm (p)}_{\rm 11} & a^{\rm (p)}_{\rm 12} \\ a^{\rm (p)}_{\rm 21} & a^{\rm (p)}_{\rm 22} 
\end {array} \right),
\eeq
which is given by
\beq
\widehat{a}^{\rm (p)}= \widehat{M}_{\rm pv}(E_{\rm f},\theta_{\rm f})\,  \widehat{a}^{\rm (v)}\, \widehat{M}^{-1}_{\rm pv}(E_{\rm i},\theta_{\rm i}),
\eeq
where the unitary matrix
\beq\label{eq:Transform_Matrix}
\widehat{M}_{\rm pv} &=& 
\frac{1}{\sqrt{1+|K_2|^2}}\left(\begin {array}{cc} -i K_2 & 1 \\ i & K_2 \end {array} \right) \nonumber \\
 &=& 
\frac{1}{\sqrt{1+|K_1|^2}}\left(\begin {array}{cc} -i &  K_1 \\ i  K_1 & 1 \end {array} \right),
\eeq
and the matrix $\widehat{a}^{\rm (v)}$ is composed of the scattering amplitudes calculated for the linearly polarised vacuum modes \citealt{1979PhRvD..19.2868H}:
\begin{align}
\label{eq:CS_Amp_01}
&a^{\rm (v)}_{\rm 11}=
\frac{E_\ii}{E_\ii+E_{\rm cyc}}e^{i(\varphi_\ii-\varphi_\ff)} 
+\frac{E_\ii}{E_\ii-E_{\rm cyc}}e^{-i(\varphi_\ii-\varphi_\ff)} \\
&a^{\rm (v)}_{\rm 22} 
= 
2\sin\theta_\ii\sin\theta_\ff\nonumber \\ 
&+ 
\cos\theta_\ii\cos\theta_\ff
\left(
\frac{E_\ii}{E_\ii+E_{\rm cyc}}e^{i(\varphi_\ii-\varphi_\ff)}
+\frac{E_\ii}{E_\ii-E_{\rm cyc}}e^{-i(\varphi_\ii-\varphi_\ff)}\right) \nonumber\\
& \left(\begin {array}{cc} a^{\rm (v)}_{\rm 12} \\ a^{\rm (v)}_{\rm 21}\end {array} \right)=
\left(\begin {array}{cc} -i\cos\theta_\ii \\ i\cos\theta_\ff\end {array} \right)\nonumber \\
&\quad\quad\quad\quad\quad
\times\left(
\frac{E_\ii}{E_\ii +E_{\rm cyc}}e^{i(\varphi_\ii-\varphi_\ff)}
-\frac{E_\ii }{E_\ii -E_{\rm cyc}}e^{-i(\varphi_\ii-\varphi_\ff)}\right).\nonumber
\end{align}

The thermal motion of electrons significantly affects both scatterings below the cyclotron resonance and at the cyclotron resonance.
\footnote{The thermal broadening of the resonance is calculated according to one-dimensional electron distribution and natural width of Landau levels approximately calculated according to \citealt{1991ApJ...380..541P}.}
In our Monte-Carlo code, we use pre-calculated tables, which describe the redistribution of photons over the energy, momentum, and polarization states.

\subsubsection{Cyclotron absorption}

Cyclotron absorption at the fundamental depends on the photon polarization state and is calculated according to \citealt{1977ewcp.book.....Z}.
In the case of $k_{\rm B}T_{\rm e}\ll m_{\rm e}c^2$,
where $k_{\rm B}$ is the Boltzmann constant, $T_{\rm e}$ is the electron temperature, and $m_{\rm e}$ is the electron mass, the adopted absorption cross section for extraordinary photons is given by
\beq
\sigma_{\rm s}^{(1)}\simeq 5.67\times 10^3 \frac{\sigma_{\rm T}}{B_{12}}\frac{(1+\cos^2\theta)}{|\cos\theta|}
\left(\frac{m_{\rm e}c^2}{k_{\rm B}T_{\rm e}}\right)^{1/2}\\
\times\exp\left[ -\frac{m_{\rm e}c^2}{2k_{\rm B}T_{\rm e}\cos^2\theta}\left(\frac{E-E_{\rm cyc}}{E_{\rm cyc}}\right)^2 \right]. 
\nonumber
\eeq
The cross section of cyclotron absorption for O-mode photons is smaller: 
$\sigma_{\rm s}^{(2)}\approx \sigma_{\rm s}^{(1)} (k_{\rm B}T_{\rm e}/m_{\rm e}c^2)$.
{In the case of $k_{\rm B}T_{\rm e}\sim m_{\rm e}c^2$ the absorption cross section is obtained numerically under the assumption that the distribution of electrons over the momentum $p_{b}$ on the ground Landau level is given by
\beq 
f_{\rm e}(p_{b})=\frac{1}{2}\frac{e^{-y\gamma}}{K_1(y)},
\eeq
where $\gamma=[1+(p_{b}/(m_{\rm e}c))^2]^{1/2}$ is the Lorentz factor, $y=m_{\rm e}c^2/(kT_{\rm e})$ is the inverse dimensionless temperature and $K_1$ is the modified Bessel function of the second kind (Mushtukov et al., in prep.).} 
Cyclotron absorption at the fundamental results in excitation of an electron from the ground Landau level to the first excited level.
Because the de-excitation rate of electrons due to emission of cyclotron photons is much larger than the de-excitation rate due to collisions between particles $r_{\rm coll}$ \citep{1979A&A....78...53B}, the majority of cyclotron absorption events are followed by almost immediate photon emission and can be considered as an event of Compton scattering (see e.g., \citealt{2006RPPh...69.2631H}).
The probability $P_{\rm abs,true}$ that the cyclotron absorption event will end up with a true absorption can be estimated from cyclotron $r_{\rm cyc}$ and collision $r_{\rm coll}$ de-excitation rates:
\beq
P_{\rm abs,true}\simeq \frac{r_{\rm coll}}{r_{\rm cyc}}\sim 1.7\times 10^{-7}n_{\rm e,21}B_{12}^{-7/2},
\eeq
where $n_{\rm e,21}=n_{\rm e}/10^{21}\,{\rm cm^{-3}}$ is local number density of electrons, which is taken in our calculations to be dependent on optical depth in the atmosphere.

\subsubsection{Free-free absorption}

Free-free (bremsstrahlung) opacity is polarization, direction and density dependent.
We calculate the opacity according to \citealt{2003ApJ...588..962L} (see Appendix \ref{App:free-free}) and under assumption that the ellipticity of normal modes is given by (\ref{eq:NormWavEll}).
The mass density $\rho$ in the atmosphere is calculated under assumption of hydrostatic equilibrium: $\d P/\d\xi=-\rho g$,
where $P$ is the gas pressure, $\xi$ - vertical coordinate in the atmosphere, and $g$ is the gravitational acceleration.
In the case of atmosphere composed of a few layers of fixed temperature, the mass density at optical depth $\tau$ is given by
\footnote{
We use the optical depth due to non-magnetic Thomson scattering, assuming that the opacity is $\kappa_{\rm T}=0.34\,{\rm cm^2\,g^{-1}}$.
}
\beq
\rho(\tau)=\rho(\tau_i)+ 40.6\,\frac{m}{T^{(i)}_{\rm keV}R_6^2} [\tau-\tau_i] \quad {\rm g\,cm^{-3}},
\eeq
where $\tau_i$ are optical depth at the top border of $i$ layer, $T^{(i)}_{\rm keV}$ is the temperature in the layer.

The initial cyclotron photons are emitted at $\sim E_{\rm cyc}$ with the thermal broadening, which depends on the direction in respect to magnetic field lines.
The angular distribution of the cyclotron photons is taken to be isotropic.

\section{Monte Carlo code}
\label{sec:MonteCarlo}

We perform Monte Carlo simulations tracing X-ray photons in plane parallel atmosphere of magnetized NS.
Magnetic field strength and its direction are fixed and assumed to be constant in the atmosphere, which is reasonable assumption because of small geometrical thickness of the atmosphere and small size of hot spots at the NS surface.
Non-linear effects of radiative transfer are not important at low luminosity states and neglected in our simulations.
Tracing X-ray photons in the atmosphere we get angular-dependent energy spectra of polarized radiation leaving the atmosphere (see Fig.\,\ref{pic:NumSch}).

{
There are two sources of seed photons in the model: the cyclotron photons due to radiative de-excitation of electrons excited to upper Landau levels by collisions of accretion flow with the atmosphere, and thermal photons.
The sources of cyclotron photons are distributed exponentially in the atmosphere. 
The cyclotron photons are emitted close to the cyclotron energy within the thermally broadened line.
The distribution of initial thermal photons is determined by the absorption coefficients in the atmosphere and local temperature (see the third term in radiative transfer equation \ref{eq:RTE}). 
The transfer of X-ray photons originated from different initial sources is considered separately.
For each simulation we trace history of $N_{\rm cyc}^{(\ii)}$ cyclotron photons and $N_{\rm th}^{(\ii)}$ thermal photons.
Because we account for free-free and cyclotron absorption in the atmosphere, the number of cyclotron and thermal photons leaving the atmosphere - $N_{\rm cyc}^{(\ff)}$ and $N_{\rm th}^{(\ff)}$ - is smaller than the number of seed photons:
$N_{\rm cyc}^{(\ff)}\le N_{\rm cyc}^{(\ii)}$ and $N_{\rm th}^{(\ff)}\le N_{\rm th}^{(\ii)}$.
Simulating radiative transfer we aim to reach a certain number ($N_{\rm th}^{(\ff)},N_{\rm cyc}^{(\ff)}\sim 2\times 10^6$) of photons that leave the atmosphere.
The results of the separately calculated radiative transfer problems are combined in a final angular and polarization dependent spectra:
\beq 
\frac{\d N_{\rm cyc}^{(\ff)}(E,\theta,s)}{\d E},
\quad\quad
\frac{\d N_{\rm th}^{(\ff)}(E,\theta,s)}{\d E}.
\eeq
In order to construct the final spectrum accounting for both sources of seed photons, we normalize the contribution of both  sources:
\beq\label{eq:sp_normalised}
\frac{\d N^{(\ff)}(E,\theta,s)}{\d E}=
A_1\,\frac{\d N_{\rm cyc}^{(\ff)}(E,\theta,s)}{\d E}+
A_2\,\frac{\d N_{\rm th}^{(\ff)}(E,\theta,s)}{\d E},
\eeq
where $A_1$ and $A_2$ are constants.
The normalization is performed on the base of the energy conservation law in the atmosphere. 
We start with the simulation of radiative transfer of the cyclotron photons.
The accretion luminosity is proportional to the total energy of seed cyclotron photons in the simulation:
\beq\label{eq:cond_1}
L_{\rm acc}\propto 
\sum_{j=1}^{N_{\rm cyc}^{\rm (i)}} E_{j,{\rm cyc}}^{\rm (i)},
\eeq 
while the part of luminosity due to thermal emission of the atmosphere is determined by the difference between total energy of seed cyclotron photons and total energy of reprocessed cyclotron photons leaving the atmosphere:
\beq \label{eq:cond_2}
\sum_{j=1}^{N_{\rm th}^{\rm (f)}} E_{j,{\rm th}}^{\rm (f)} =
\left[\sum_{j=1}^{N_{\rm cyc}^{\rm (i)}} E_{j,{\rm cyc}}^{\rm (i)} - 
\sum_{j=1}^{N_{\rm cyc}^{\rm (f)}} E_{j,{\rm cyc}}^{\rm (f)} \right].
\eeq 
Using conditions (\ref{eq:cond_1}) and (\ref{eq:cond_2}) we get constants $A_1$ and $A_2$ in (\ref{eq:sp_normalised}).
}

{
In order to perform Monte Carlo simulations and track the photons  we use a set of pre-calculated tables describing photon redistribution due to magnetic Compton scatterings:
}

{
{\it Tables A} of total scattering cross-section, where the cross-section is given as a function of photon energy, polarisation state before and after the scattering event, the angle between the $B$-field direction and photon momentum, temperature and bulk velocity of the electron gas.
For each combination of the initial and final polarization state of a photon, the tables are pre-calculated in a fixed grid in photon energy and angle $\theta_{\rm i}$, and for a fixed temperature and bulk velocity of the gas.
To get a scattering cross-section for given photon energy and momentum, we use quadratic interpolation in the photon energy grid and further quadratic interpolation in the angle grid.
}

{
{\it Tables B} of photon probabilities to be scattered into a certain segment of the solid angle $(\theta_{\rm f}+\Delta\theta_{\rm f},\varphi_{\rm f}+\Delta\varphi_{\rm f})$.
The tables are pre-calculated on a grid of photon initial parameters (energy, polarization state, momentum) and for both possible final polarization states. 
}

\smallskip\smallskip
{
Steps of tracing of photon history are
\begin{enumerate}
\item\label{ref:MC_start} We make a choice of seed photon origin: thermal emission of the atmosphere or cyclotron emission.
If the photon is due to the cyclotron emission, we get the optical depth where the photon is emitted:
$$\tau_0=-\tau_{\rm br}\ln X_1,$$
where $X_1\in(0;1)$ is a random number,
and the photon energy assuming that the photon is emitted within a thermally broadened cyclotron line.
If the photon is due to the thermal emission, we get the optical depth of its emission in the atmosphere out of pre-calculated cumulative distribution functions of photon emission accounting for the assumed temperature structure in the atmosphere and free-free absorption coefficient.
\item\label{step:get_free_path} We get the free path of the photon accounting for scattering and absorption opacity.
\item 
Using the initial coordinate of the photon and free path length, we get a new coordinate of a photon, where it is scattered or absorbed.
If the new coordinate is located out of the atmosphere, the photon contributes to the spectra of X-ray radiation leaving the atmosphere, and we come back to the step \ref{ref:MC_start}.
If the photon is still in the atmosphere, we move on to step \ref{step:choose_process}.
\item\label{step:choose_process}
Comparing the cross section of Compton scattering, free-free and cyclotron absorption, we specify the elementary process at a new photon coordinate.
If the photon is absorbed, we stop trace history of the photon and start tracing a new photon, i.e. come back to step \ref{ref:MC_start}. 
If the photon is scattered by electrons, we get its new energy, momentum direction and polarization state on the base of pre-calculated tables, and come back to step \ref{step:get_free_path}.
\end{enumerate}
}

\section{Results of numerical simulations}
\label{sec:NumRes}

\subsection{Influence of NS atmosphere conditions on the X-ray spectra} 

For the case of fixed local magnetic field strength, there are five main parameters of the performed numerical simulations and resulting spectra of X-ray photons. 
These are 
(i) the optical thickness of the overheated upper layer due to Thomson scattering $\tau_{\rm up}$,
(ii) the typical length of accretion flow braking in the atmosphere measured in optical depth due to Thomson scattering $\tau_{\rm br}$,
(iii) the temperature of the atmosphere under the overheated upper layer $T_{\rm bot}$, 
(iv) the temperature of the overheated upper layer $T_{\rm up}$,
and (v) the chemical composition of the atmosphere given by the atomic number $Z$.
Here we investigate how these parameters affect the spectrum.
To separate effects caused by different reasons, we compare the results of numerical simulations with the results based on the following set of fiducial parameters:
$T_{\rm bot}=3\,{\rm keV}$, 
$T_{\rm up}=80\,{\rm keV}$, 
$\tau_{\rm up}=0.1$, 
$\tau_{\rm br}=1$, 
$Z=1$, $\theta_{\rm B}=0$ (red line in Fig.\,\ref{pic:MC_comparison}\,a-f).

In most cases, we see the X-ray energy spectrum consisting of two components. 
The low energy component is a result of black-body radiation comptonized by electrons in the atmosphere. 
The high energy component is a result of the initial emission of cyclotron photons and their further comptonization by electrons. 
The multiple scatterings of cyclotron photons are strongly affected by the resonance at the cyclotron energy broadened by the thermal motion.
X-ray photons hardly escape the atmosphere at the energies close to the cyclotron energy, and because of that, the photons tent to escape in red and blue wings of a cyclotron line.
Note that thermal emission also contributes to the initial photons at cyclotron energy because free-free absorption is resonant at $E_{\rm cyc}$ in X-mode. 

Comparing the results of different numerical simulations, we can make some conclusions:\\
{
(a) The larger optical thickness of the overheated upper layer $\tau_{\rm up}$ does not affect much the low-energy part of X-ray spectra, but influence the high energy component (see Fig.\,\ref{pic:MC_comparison}a) affecting photon redistribution around the cyclotron line.\\ 
(b) Smaller optical depth of accretion flow braking $\tau_{\rm br}$ leads to stronger high-energy component of X-ray spectrum (see Fig.\,\ref{pic:MC_comparison}b). 
It is natural because at smaller $\tau_{\rm br}$ it becomes easier for cyclotron photons to leave the atmosphere, starting their diffusion from a smaller optical depth. 
Additionally, the smaller optical depth of the accretion flow braking results in a smaller fraction of absorbed cyclotron photons, contributing to the thermal low-energy part of X-ray spectra.\\
(c) The overheated upper layer of the atmosphere affects the high-energy end of X-ray spectra.
The lower temperature of the upper layer $T_{\rm up}$ makes the blue wing of a cyclotron line weaker (see Fig.\,\ref{pic:MC_comparison}c). If the temperature of the upper layer is much smaller than the cyclotron energy $E_{\rm cyc}$, the most of cyclotron photons are scattered into a red wing of the line and leave the atmosphere at $E\lesssim E_{\rm cyc}$.\\
(d) The temperature $T_{\rm bot}$ of the atmosphere below the overheated upper layer shapes the thermal radiation of the atmosphere and
affects the low energy part of X-ray spectra.
Increase of the temperature of the bottom atmosphere results in a corresponding shift of low-energy component (see Fig.\,\ref{pic:MC_comparison}d).\\
(e) The chemical composition in the atmosphere affects the cross section of free-free absorption: the larger the atomic number, the larger the cross section of free-free absorption.
Because the low-energy component is dominated by the thermal emission of the atmosphere, this component is affected stronger by the variations in free-free absorption coefficient. 
Specifically, we see that the energy spectra tend to be slightly suppressed at low energies for larger effective atomic numbers $Z$ (see Fig.\,\ref{pic:MC_comparison}e).\\
(f) The direction of magnetic field in respect to the NS surface does not affect much the final spectra integrated over the solid angle (see Fig.\,\ref{pic:MC_comparison}f). In our simulations, we see only a slight decrease of the width of the absorption feature at $E\sim E_{\rm cyc}$.
This decrease is likely due to the fact that the thermal broadening of the cyclotron resonances in Compton scattering cross section is weaker for the photons propagating across the field direction.}\\

\begin{figure}
\centering 
\includegraphics[width=8.5cm]{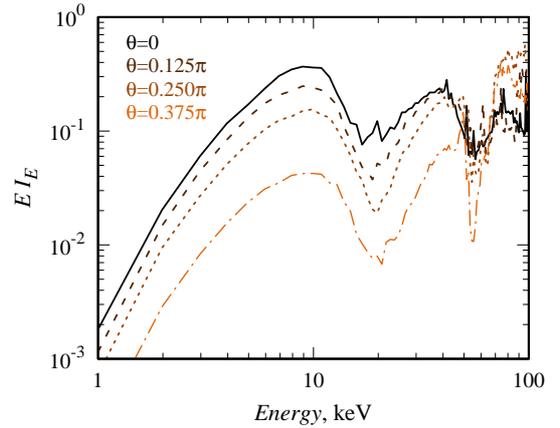} 
\caption{
{Specific intensity $I_E$ at the stellar surface at different directions given by an angle $\theta$ between local $B$-field direction and photon momentum. 
Different lines are given for different angles $\theta=0$ (solid),
$0.125\pi$ (dashed),
$0.25\pi$ (dotted),
$0.375\pi$ (dashed-dotted).
Note, that the intensity in red and blue wings of a cyclotron line are strongly variable. 
Parameters for simulated spectrum: $E_{\rm cyc}=50\,{\rm keV}$, $T_{\rm bot}=3\,{\rm keV}$, $T_{\rm up}=80\,{\rm keV}$, $\tau_{\rm up}=0.1$, $\tau_{\rm br}=1$, $Z=1$.}
}
\label{pic:MC_angle}
\end{figure}

The specific intensity of X-ray radiation leaving the atmosphere is angular dependent (see Fig.\,\ref{pic:MC_angle}).
In particular, a strong angular dependence of the specific intensity at red and blue wings of a cyclotron line is expected.
The intensity integrated over the energies composes a pencil beamed diagram, which is slightly suppressed in the direction orthogonal to the stellar surface.
This phenomenon is natural for the case of the atmosphere with the overheated upper layer: the contribution of the overheated upper layer into the intensity is smaller in the direction perpendicular to the stellar surface. 

\begin{figure}
\centering 
\includegraphics[width=8.5cm]{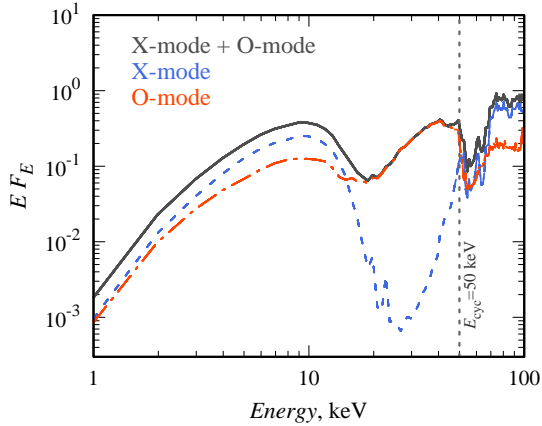} 
\caption{
X-ray energy spectra at the NS surface in X-mode (blue dashed line) and O-mode (red dashed-dotted line) and complete spectra (black solid line). 
Parameters for simulated spectrum: $E_{\rm cyc}=50\,{\rm keV}$, $T_{\rm bot}=3\,{\rm keV}$, $T_{\rm up}=80\,{\rm keV}$, $\tau_{\rm up}=0.1$, $\tau_{\rm br}=1$, $Z=1$.
}
\label{pic:MC_modes}
\end{figure}

The X-ray energy flux leaving the atmosphere is polarization-dependent (see Fig.\,\ref{pic:MC_modes}).
The polarization dependence is particularly strong near the cyclotron energy, which is natural because the strength of cyclotron resonance and even its existence depends on the polarization state of a photon.
At low energies $E\ll E_{\rm cyc}$ the flux is dominated by X-mode photons because the scattering cross-section is smaller for this polarization state \citep{1979PhRvD..19.2868H,1986ApJ...309..362D,2016PhRvD..93j5003M}.
However, the difference in X-ray energy flux at different polarization states is not dramatic because of the inverse temperature profile in the atmosphere: the upper layer is assumed to be much hotter than the underling atmosphere.
Note, however, that the exact predictions for polarization require detailed analyses of effects arising from vacuum polarization and complicated behavior of plasma dielectric tensor under conditions of high temperatures. 
The detailed analysis of these effects is beyond the scope of this paper and will be discussed in a separate publication.

\subsection{Comparison to the observational data}
\label{sec:obs}

In order to verify our model, we compared results of the simulations with the data obtained during a low-luminosity state of transient XRP A~0535+262 with $L_{\rm X}=7\times10^{34}$~\lum. 
The data were adopted from \cite{2019MNRAS.487L..30T} and cover broad-energy band from 0.3 to 79 keV using {\it Swift}/XRT and {\it NuSTAR} instruments.

\begin{figure}
\centering 
\includegraphics[width=8.7cm]{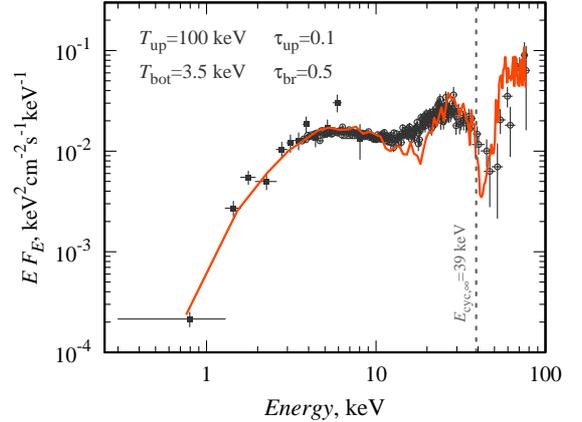} 
\caption{
Observed $E\,F_E$ spectrum of A~0535+262 at accretion luminosity of $~7\times 10^{34}\,\ergs$ is given by black circles (\textit{NuSTAR} FPMA and FPMB data) and squares (\textit{Swift}/XRT). 
Simulated spectrum is represented by red line.
Parameters for simulated spectrum: $E^{\infty}_{\rm cyc}=39\,{\rm keV}$, $T_{\rm bot}=2.8\,{\rm keV}$, $T_{\rm up}=100\,{\rm keV}$, $\tau_{\rm up}=0.1$, $\tau_{\rm br}=0.5$, $Z=1$.
The run results from $5\times 10^5$ photons leaving the atmosphere.
}
\label{pic:MC_vs_Data}
\end{figure}

{
In Fig.~\ref{pic:MC_vs_Data} we represent the model calculated for the following set of parameters (see red line): 
\beq
&E_{\rm cyc}=48\,{\rm keV},\quad\quad\quad 
Z=1,\quad\quad\quad
\theta_{\rm B}=0  \nonumber \\
&T_{\rm bot}=3.5\,{\rm keV},\quad\quad\quad
T_{\rm up}=100\,{\rm keV},&\nonumber \\ 
&\tau_{\rm up}=0.1,\quad\quad\quad
\tau_{\rm br}=0.5.&\nonumber 
\eeq
The theoretical model represents the spectrum integrated over the solid angle at the NS surface and is based on Monte Carlo simulation with $2\times 10^6$ photons, leaving the atmosphere. 
In order to compare the theoretical model with the observed X-ray spectra, we have accounted for spectral changes due to the gravitational red shift. 
We assume that the photon energy detected by a distant observer $E^{\infty}$ is related to the photon energy at the NS surface $E$ as 
\beq 
E_{\infty}=E(1-u)^{1/2},
\eeq 
where $u=R_{\rm S}/R$, and Schwarzschild radius $R_{\rm S}\approx 3m\,{\rm km}$.
Mass and radius of a NS were taken to be $M=1.4\,M_\odot$ and $R=12\,{\rm km}$.
As can be seen, our theoretical predictions are able to describe complex spectral shape of the source including all observed features.
}

We note, however, that our theoretical model shows a lack
of X-ray photons in a red wing of a cyclotron line at energy $\sim 15-20\,{\rm keV}$.
It is hard to eliminate this discrepancy in the energy spectrum integrated over the solid angle.
We suppose that this problem can be solved if one accounts for the exact geometry of NS rotation in the observer's reference frame and the precise process of pulse profile formation.
We also note that the radiative transfer at the high energy part of X-ray spectra can be affected by the effect of vacuum polarisation (see Appendix \ref{App:vac_pol}), which was not taken into account in our simulations. 
This analysis is beyond the scope of the paper and a matter of further investigations, which will be discussed in a separate publication.

\section{Summary and Discussion}
\label{sec:sum}

\subsection{Spectra formation at low mass accretion rates}

We performed numerical simulations for spectra formation in XRPs at very low-luminosity state, when the interaction between radiation and accretion flow above NS surface does not affect X-ray spectra and dynamics of the accretion flow.
Our simulations are coherent with a physical model (see Fig.\,\ref{pic:000}) where the accretion flow is braked in upper layers of NS atmosphere due to collisions between particles, and most of kinetic energy is released initially in the form of cyclotron photons. The spectra leaving the atmosphere of a NS is a matter of radiative transfer, strongly affected by magnetic Compton scattering. 
The essential ingredient of the model is an overheated upper layer of the NS atmosphere proposed earlier by \citealt{2018A&A...619A.114S} for the case of low-level accretion onto weakly magnetized NSs.
Simulated radiative transfer in the atmosphere was performed under the assumption of LTE and accounts for Compton scattering of X-ray photons by thermally distributed electrons, cyclotron, and free-free absorption.
Two components in the spectrum correspond to comptonized thermal radiation (low-energy hump), and comptonized cyclotron photons (high-energy hump) originated from collisions of accreting particles with the electrons in the NS atmosphere and further radiative transition of electrons to the ground Landau level.
The absorption feature on top of the high-energy hump is due to the resonant scattering of X-ray photons at cyclotron energy, which forces cyclotron photons to leave the atmosphere in the wings of a cyclotron line.

Using the constructed model, it was possible to reproduce the observed spectrum of X-ray pulsar A~0535+262 at a very low luminosity state (see Fig.\,\ref{pic:MC_vs_Data}, \citealt{2019MNRAS.487L..30T}).
Qualitative agreement between simulated and observed X-ray spectra confirms assumptions of the underlying physical model. 
We argue that two-component spectra should be typical for low-level accretion onto strongly magnetized NSs. 

\subsection{Applications to the observational studies}

\subsubsection{Investigation of the ``propeller" state}

Decreasing of the mass accretion rates in XRPs down to the very low values results in the transition of the source either to (i) the ``propeller" state, when the accretion flow cannot penetrate through the centrifugal barrier set up by the rotating magnetosphere of a NS (see, e.g., \citealt{1975A&A....39..185I,1987Ap&SS.132....1L,2006ApJ...646..304U}), or (ii) to the regime of stable accretion from a cold disc (see \citealt{2017A&A...608A..17T}). 
``Propeller" effect was recently detected in a few XRPs \citep[see e.g.,][]{2016A&A...593A..16T,2017ApJ...834..209L}.
Moreover, in some sources, transitions into the ``propeller" state were discovered to be accompanied by dramatic changes of X-ray energy spectra \citep{2016A&A...593A..16T}. 
Specifically, in the energy range below $\sim10$~keV, the spectra were shown to become significantly softer with shape changed from the power-law to a black body with a typical temperature around 0.5 keV. 
However, it is still unknown if the centrifugal barrier blocks the accretion process entirely, and the detected soft X-ray spectra are observational evidence of cooling NS surface, or leakage of matter through the barrier is still possible and responsible for some fraction of the observed emission \citep{2016MNRAS.463L..46W,2017MNRAS.472.1802R}.

Our theoretical model of spectra formation provides a natural way to distinguish low-level accretion from the cooling NS surface.
The hard component of X-ray spectra is a direct result and a specific feature of the accretion process.
As a result, low-level accretion in the case of the leakage of the centrifugal barrier should result in two-component X-ray spectra, 
while the ``propeller" state without penetration of material through the barrier should result in single hump soft spectra.

\subsubsection{Measurements of magnetic field strength}

Two-component X-ray energy spectra at low mass accretion rate provide a way to estimate magnetic field strength at the NS surface.
Indeed, the hard component of X-ray spectra is formed around cyclotron energy, which is directly related to the field strength: $E_{\rm cyc}\approx 11.6\,B_{12}\,$keV.
Thus, the detection of hard-energy hump provides a way to estimate the cyclotron energy in the case when cyclotron absorption feature is not seen in the source spectrum. 

For instance, the spectra of X Persei \citep{1998ApJ...509..897D}, according to our model, implies the cyclotron energy $E_{\rm cyc}\gtrsim 100\,{\rm keV}$ and corresponding magnetic field strength $B\gtrsim 10^{13}\,{\rm G}$. 
Note that this estimation is coherent with the results based on timing analyses and torque models applied to this particular source \citep{2012A&A...540L...1D}.

\section*{Acknowledgements}

This work was supported by the Netherlands Organization for Scientific Research Veni Fellowship (AAM), the V\"ais\"al\"a Foundation (SST) and the Academy of Finland travel grant 324550. VFS thanks Deutsche  Forschungsgemeinschaft  (DFG) for financial support (grant WE 1312/51-1).
The authors also thank the Russian Science Foundation (grant 19-12-00423) for financial support.
We are grateful to an anonymous referee for a number of useful comments and suggestions which helped us improve the paper.

\section*{Data availability}

The calculations presented in this paper were performed using a private code developed and owned by the corresponding author, please contact him for any request/question about. Data appearing in the figures are available upon request. 
The observational data used in the manuscript are adopted from those reported in \citealt{2019MNRAS.487L..30T}.

\bibliographystyle{mnras}

\bibliography{Low_State_Spec}

\appendix
\section{Free-free absorption in a strong magnetic field}
\label{App:free-free}

The opacity due to free-free absorption is calculated in our simulations according to \citealt{2003ApJ...588..962L}.
The opacity is dependent on magnetic field strength, polarization of X-ray photons, their energy and direction of the momentum in respect to local direction of $B$-field.
The opacity for mode $j$ of ellipticity $K_j$ can be written as 
\beq\label{eq:kappa_opas}
\kappa_j^{\rm ff}=\kappa_{+}|e_{+}^j|^2 + \kappa_{-}|e_{-}^j|^2 + \kappa_{0}|e_{0}^j|^2,
\eeq
where $e_0$, $e_\pm$ are the spherical components of the photon's unit polarization vector with
\beq
|e_{\pm}^j|^2 &=& \left| \frac{1}{\sqrt{2}}(e_x^j \pm i e_y^j) \right|^2 \\
&=&\frac{[1\pm (K_j\cos\theta + K_{z,j}\sin\theta)]^2}{2(1+K_j^2+K_{z,j}^2)}, \nonumber \\
|e_z^j|^2 &=& \frac{(K_j\sin\theta - K_{z,j}\cos\theta)^2}{1+K_j^2+K_{z,j}^2},
\eeq
where the ellipticities $K_j$ are given by (\ref{eq:NormWavEll}) and $K_{z,j}$ is taken to be $0$.
The coefficients in (\ref{eq:kappa_opas}) are given by
\beq
\kappa_\pm=\frac{\omega}{c\rho}v_e\Lambda_\pm\gamma_{ei}^{\perp},\quad\quad
\kappa_0=\frac{\omega}{c\rho}v_e \gamma_{ie}^{\parallel},
\eeq
where
\beq
\Lambda_\pm = [(1\pm u_e^{1/2})^2(1\mp u_i^{1/2})^2+\gamma_\pm^2]^{-1}, \\
\gamma_\pm = \gamma_{ei}^\perp + (1\pm u_e^{1/2})\gamma_{ri} + (1\mp u_i^{1/2})\gamma_{re}.
\eeq
Dimensionless quantities:
\beq
u_e=\left(\frac{E_{\rm cyc}}{E}\right)^2,\quad
u_i=\left(\frac{E_{Bi}}{E}\right)^2,\quad
v_e=\left(\frac{E_{pe}}{E}\right)^2,
\eeq
where the energies corresponding to electron cyclotron frequency, ion cyclotron frequency, and the electron plasma frequency are given by
\beq
E_{\rm cyc}&=&11.58\,B_{12}\,{\rm keV}, \\
E_{Bi}&=&6.305\times 10^{-3}\,B_{12}\left(\frac{Z}{A}\right)\,{\rm keV}, \\
E_{pe}&=&2.871 \times 10^{-2}\left(\frac{Z}{A}\right)^{1/2}\rho^{1/2}\,{\rm keV}.
\eeq
The dimensionless dumping rates due to electron-ion collisions and photon emission by electrons and ions are given by
\beq
\gamma_{ei}^{\perp,\parallel} &=& 4.55\times 10^{-8}\,\frac{Z^2 n_{i,21}}{T_{\rm keV}^{1/2}E_{\rm keV}^{2}}
\left(1-e^{-E/k_{\rm B}T}\right) g^{\rm ff}_{\perp,\parallel}, \nonumber \\
\gamma_{re}&=&9.52\times 10^{-6}\,E_{\rm keV}, \\
\gamma_{ri}&=& 7.1\times 10^{-7}\,\frac{Z^2}{A}\,E_{\rm keV}, \nonumber
\eeq
where $g^{\rm ff}_{\perp,\parallel}$ are magnetic Gaunt factors, which are calculated according to \citealt{2010ApJ...714..630S} in our simulations.

\section{On the influence of vacuum polarization}
\label{App:vac_pol}

Assuming constant temperature $T$ in the atmosphere one can get the dependence of the mass density on the vertical coordinate $\xi$:
\beq
\rho(\xi)=\rho(\xi_0)\exp\left[ \frac{g m_{\rm p}}{k_{\rm B}T}(\xi-\xi_0) \right], 
\eeq
where $g\simeq 1.3\times 10^{14}\,m R_6^{-2}\,{\rm cm\,s^{-2}}$ is the acceleration due to gravity at the NS surface.
If the opacity is given by Thomson opacity $\kappa_{\rm e}=0.34\,{\rm cm^2\,g^{-1}}$, the local mass density is related to the optical depth in the atmosphere as
\beq
\label{eq:rho_a}
\rho=40.6\,\tau_{\rm T}\frac{m}{T_{\rm keV}R_6^2}\quad {\rm g\,cm^{-3}}.
\eeq

The critical mass density, when the contribution of vacuum polarisation becomes comparable to the contribution of plasma to the dielectric tensor, can be estimated as \citep{2003ApJ...588..962L}
\beq
\label{eq:rho_v}
\rho_{\rm V}=9.64\times 10^{-5}\,Y_{\rm e}^{-1}B_{12}^2 E_{\rm keV}^2 f^{-2}\quad {\rm g\,cm^{-3}},
\eeq
where $Y_{\rm e}=Z/A$ and $f\sim 1$.
At $\rho\ll \rho_{\rm V}$ the dielectric tensor is dominated by vacuum effects, while
at $\rho\gg \rho_{\rm V}$, the dielectric tensor is dominated by plasma effects.
Using (\ref{eq:rho_a}) and (\ref{eq:rho_v}) we get the optical depth due to Thomson scattering corresponding to $\rho_{\rm V}$:
\beq
\tau_{\rm T,V}=2.4\times 10^{-6}\,Y_{\rm e}^{-1}B_{12}^2 T_{\rm keV} E_{\rm keV}^2 f^{-2} m^{-1} R_6^2. 
\eeq
Therefore, for the case of physical conditions expected in XRP A~0535+262 we get
$$
\tau_{\rm T,V}\sim 3\times 10^{-5}T_{\rm keV}E_{\rm keV}^2,
$$
which means that the polarization of X-ray photons in low-energy hump is well described by (\ref{eq:NormWavEll}), while polarization of X-ray photons in high-energy hump might be affected by the effects of vacuum polarization, which will be investigated in a separate publication.

\bsp 
\label{lastpage}
\end{document}